\documentclass[pra,twocolumn,nofootinbib,notitlepage]{revtex4-1}%twocolumn,
\usepackage{graphicx}% Include figure files
\usepackage{bm}% bold math
\usepackage{color}
\usepackage{amsmath, mathtools}
\usepackage{amssymb}
\usepackage{wasysym}
\usepackage{booktabs} % ``Proper'' table layout
\usepackage{pdfsync}
\usepackage{verbatim}
\usepackage{latexsym}
\usepackage{dsfont}
\def\sec#1{\section{#1} }
\def\ssec#1{\subsection{#1} }

\def\({\left(}
\def\){\right)}
\def\[{\left[}
\def\]{\right]}

\def\a{\alpha}

\def\f#1#2{\frac{#1}{#2}}
\def\g{\gamma}

\def\d{\partial}
\def\de{\delta}

\def\del{\nabla}

\def\vep{\varepsilon}

\def\k{\kappa}

\def\m{\mu}

\def\o{\omega}

\def\p{\pi}

\def\s{\sigma}

\def\th{\theta}

\def\MeV{\text{MeV}}

\def\<{\langle}
\def\>{\rangle}

\providecommand{\abs}[1]{\left\lvert#1\right\rvert}

\begin{document}

\title{An artificial boundary approach for short-ranged interactions}% non-relativistic quantum mechanics

\author{David M. Jacobs}
\email{dmj15@case.edu}
\affiliation{Astrophysics, Cosmology and Gravity Centre,\\
Department of Mathematics and Applied Mathematics,\\
University of Cape Town\\
Rondebosch 7701, Cape Town, South Africa}

%\maketitle
%\abstract{
\begin{abstract}
%Boundary conditions of wavefunctions and fields can be used to model short-range interactions when those interactions are expected, \emph{a priori}. 
Real physical systems are only understood, experimentally or theoretically, to a finite resolution so in their analysis there is generally an ignorance of possible short-range phenomena. It is also well-known that the boundary conditions of wavefunctions and fields can be used to model short-range interactions when those interactions are expected, \emph{a priori}. Here, a real-space approach is described wherein an artificial boundary of ignorance is imposed to explicitly exclude from analysis the region of a system wherein short-distance effects may be obscure.
%It has been known for some time that the boundary conditions of wavefunctions and fields can be used to model short-range interactions when those interactions are expected, \emph{a priori}. But real physical systems are only understood, experimentally or theoretically, to a finite resolution so in their analysis there is generally an ignorance of possible short-range phenomena. To address this, a real-space approach is described wherein an artificial boundary of ignorance is imposed to explicitly exclude from analysis the region of a system wherein short-distance effects may be obscure.
 The (artificial) boundary conditions encode those short-distance effects by parameterizing the possible UV completions of the wavefunction. Since measurable quantities, such as spectra and cross sections, must be independent of the position of the artificial boundary, the boundary conditions must evolve with the radius of the boundary in a particular way.  As examples of this approach, an analysis is performed of the non-relativistic free particle, harmonic oscillator, and Coulomb potential, and some known results for point-like (contact) interactions are recovered, however from a novel perspective. 
Generically, observables differ from their canonical values and symmetries are anomalously broken compared to those of idealized models.  Connections are made to well-studied physical systems, such as the binding of light nuclei and cold atomic systems.  This method is arguably more physically transparent and mathematically easier to use than other techniques that require the regularization and renormalization of delta-function potentials, and may offer further generalizations of practical use.
\end{abstract}

\maketitle

\sec{Introduction}
%{\bf Introduction}:
The study of boundary conditions for quantum mechanical wavefunctions has a long history that dates back to Weyl and von Neumann (see e.g. \cite{Bonneau:1999zq}). It has been appreciated, although perhaps not widely, that boundary conditions generically describe short-ranged interactions that are not explicitly modeled by interaction terms in the Hamiltonian. Although the arguments one may find in introductory textbooks on the subject can lead one to believe that there is always a unique choice of boundary condition, that choice often comes from a set of many physically-acceptable alternatives -- this falls under the subject of self-adjoint extensions\footnote{A well-known fundamental tenet of quantum mechanics is that observables are represented by self-adjoint operators, i.e. Hermitean operators whose domain is equal to that of its adjoint.}.
%\footnote{For example, a common argument in favor of choosing a vanishing wavefunction is that the supposed continuity of the wavefunction; this, however, is a fallacious arugment that does not apply at a true boundary.}
 %One motivation of the present work is to advertise the fact that using non-trivial boundary conditions can extend the range of utility of a theory by accounting for short-range effects in an effective way that are known about, yet sufficiently complicated, or perhaps even to capture an unanticipated effect.
%A related issue is the fact that theories can only be trusted to finite distance scales due to either a known limit to their range of validity and/or a lack of experimental confirmation at smaller scales. Specifying the boundary condition of a wavefunction, however, suggests having infinitely precise knowledge of what the wavefunction does in an infinitesimally small region of space. Surely, such infinite precision isn't needed to describe nature to finite accuracy. 
Self-adjoint extensions have been applied in pedagogical inquiries  \cite{Bonneau:1999zq, essin2006quantum, Dasarathy:unpub}, as an alternative to delta function potentials in two or three dimensions\footnote{The difficulties in using a delta function potential in two or three dimensions have long been appreciated. Techniques from field theory, namely regularization and renormalization must be used for that approach to work and it has been advocated that self-adjoint extensions are an easier, but equivalent method \cite{jackiw1995diverse}.} \cite{jackiw1995diverse}, and, similarly, to model point-like (or contact) interactions (e.g. \cite{Coutinho, Roy:2009vc, albeverio2012solvable}).

Recently, the non-relativistic Coulomb problem has been revisted using self-adjoint extensions \cite{Beck:inprep}. In that work, it is shown that boundary conditions can be used to describe the type of contact interactions that are usually approximated by delta functions such as the Darwin fine-structure correction. Using self-adjoint extensions, a non-trivial boundary condition is be imposed on the $\ell=0$ Coulomb wavefunction at the origin and can uniquely chosen by matching the resulting modified spectrum to the known fine-structure spectrum.  A dimensionless measure of the deviation of the boundary condition from the standard choice is $\a^2$, where $\a\simeq 1/137$ is the fine structure constant.  This number being small suggests one reason why this approach had not been appreciated before.

The index theorem discovered by Weyl, and latter generalized by von Neumann, is used to identify how many self-adjoint extension parameters a given Hamiltonian has \cite{Bonneau:1999zq}.  For some systems the Hamiltonian is \emph{essentially self-adjoint}, meaning there is a unique self-adjoint extension that yields a well-defined theory. However, in general, there can exist a multi-dimensional family of boundary conditions that specify the domain of a Hamiltonian, with no compelling physical argument to choose any particular set, \emph{a priori}\footnote{Singular or discontinuous wavefunctions are not necessarily problematic if the singularity or discontinuity occurs at a boundary.}.

Some boundary conditions are special as they ensure that a system possesses a particular symmetry such as $SO(4)$ for the case of bound Coulomb states \cite{Beck:inprep}. However, it is also well-known that classical symmetries can be destroyed by quantum effects, so all extensions should be given consideration. This is a philosophy already appreciated in effective field theory, where \emph{all} possible operators must be included in the theory's action that are not forbidden by some symmetry principle; comparison to experimental data or a deeper underlying theory is required to fix the coefficients of those operators.

%At any rate, the insistence on certain boundary conditions, such as the vanishing of the wavefunction at the boundary, misses a wealth of other possibilities that correspond to \emph{real} physical systems, as noted in \cite{Bonneau:1999zq}. 

The goals of this article are three-fold: to emphasize that boundary condition freedom (self-adjoint extensions in quantum mechanics) should be exploited for their maximum utility in \emph{all} systems because they effectively capture short-range interactions that may be present; to describe an artificial boundary method for identifying those freedoms and computing observables; and to make connections between this work, established results for self-adjoint extensions, and previously-studied models that describe short-range interactions, such as those found in ultra-cold atomic systems \cite{chin2010feshbach}.

Although many results presented in this work may be unfamiliar to the reader, no prediction made here is fundamentally new since they have been derived elsewhere, although perhaps in seemingly unrelated contexts. One motivation for the present work is to provide a unified framework in which to arrive at those results, a framework that may be useful for studying other systems.   Another motivation is the hope that looking at old problems from a different point of view may provide new insights for solving outstanding ones (see e.g. \cite{Feynman:1948ur}).

The guiding physical principle here is the conservation of probability which, in the field-theoretic context, corresponds to global conservation of energy -- both of these follow from a system's invariance under time translations. With a standard textbook treatment it can be shown that the time independence of the norm of the wavefunction,
\begin{equation}
(\Psi,\Psi)\,,
\end{equation}
requires
\begin{equation}\label{prob_consv_condition}
(\Psi,H\Psi)=(H\Psi,\Psi)\,,
\end{equation}
where $H$ is the Hamiltonian, and follows from the Schr\"odinger equation. For this to hold for any wavefunction actually requires
\begin{equation}\label{hermitean_condtion}
(\Phi,H\Psi)=(H\Phi,\Psi)\,,
\end{equation}
for all $\Psi$ and $\Phi$ in the domain of $H$ -- this ensures that $H$ is a Hermitean operator; to further restrict the domain of $H^\dagger$ to be the same as $H$ ensures it is self-adjoint \cite{reed1980methods}. For the systems analyzed here, imposing the most general boundary condition that satisfies \eqref{prob_consv_condition} is sufficient to guarantee the Hamiltonian is self-adjoint.

\sec{Preliminaries for spherically-symmetric systems}

For purposes of illustration, the Hamiltonians considered in this work are of the spherically symmetric form\footnote{Throughout, units are chosen wherein $\hbar=1$ unless otherwise specified.}
\begin{equation}
H=-\f{1}{2\m} \del^2 +V(r)\,,
\end{equation}
where $\m$ refers to either a single particle mass or reduced mass of a two-particle system.

In order to effectively capture any short-distance interactions near the origin, an artificial boundary is placed at $r=r_\star$ so that the region $0\leq r<r_\star$ is not part of the configuration space. Despite the apparent violence this does to the original model, if analysis is restricted to modes of wavenumbers, $k$, satisfying the condition $k r_\star\ll 1$ there is plausibly no significant error by inserting a boundary in this fashion since particles cannot be localized to a resolution of the order $r_\star$. Alternatively, this can be seen as a low-energy limit in a theory with a cutoff scale of $1/(2m r_\star^2).$ In any case, \emph{the limit $r_\star\to0$ will be taken at the end of the calculation in order to recover the original domain.} 

If the wavefunction and its derivatives vanish sufficiently rapidly as $r\to\infty$, then to satisfy \eqref{prob_consv_condition} the boundary integral
\begin{equation}\label{surface_integral}
r_\star^2\int d\Omega \, \[\bar{\Psi} \d_r\Psi - \(\d_r\bar{\Psi}\) \Psi  \]_{r=r_\star}
\end{equation}
must vanish. Because of the spherical symmetry, the usual ansatz is made for a particular eigenmode,
\begin{equation}
f_{\ell}(r) Y_{\ell m}(\Omega)\,,
\end{equation}
where $Y_{\ell m}(\Omega)$ is a properly normalized spherical harmonic. Since \eqref{surface_integral} must vanish for arbitrary $r_\star$, it must be that
\begin{equation}\label{3d_boundary_term}
\bar{f}_\ell(r_\star) f_\ell'(r_\star) - \bar{f}'_\ell(r_\star) f_\ell(r_\star) =0\,.
\end{equation}
Following \cite{Bonneau:1999zq}, the identity
\begin{equation}
\(x\bar{y}-\bar{x}y\)=\f{i}{2}\(\abs{x+iy}^2-\abs{x-iy}^2\)
\end{equation}
may be used to rewrite the condition \eqref{3d_boundary_term} as
\begin{equation}\label{new_condition}
\abs{f_\ell(r_\star)+iw_\ell f'_\ell(r_\star)}^2 - \abs{f_\ell(r_\star)-iw_\ell f'_\ell(r_\star)}^2=0\,,
\end{equation}
where the $w_\ell$ are arbitrary real-valued constants with units of length and are only inserted for dimensional reasons. The two terms in \eqref{new_condition} must be equal up to a phase, $e^{-i\chi_\ell}$, where $0 \leq\chi_\ell <2\p$. It follows that the general boundary condition is
\begin{equation}\label{spher_symm_SAE_condition}
f_\ell(r_\star) + Z_{\ell}(r_\star)\,f'_\ell(r_\star)=0\,,
\end{equation}
where the function $Z_\ell(r_\star)\equiv w_\ell\cot{\f{\chi_\ell}{2}}$  can take any real value. It is straightforward to show that \eqref{hermitean_condtion} can only hold if all radial eigenfunctions obey \eqref{spher_symm_SAE_condition}. Lastly, restricting $H^\dagger$ to the domain defined by \eqref{spher_symm_SAE_condition} ensures that $H$ is self-adjoint. One then needs to determine the $Z_{\ell}(r_\star)$ for a given potential.

\sec{``Free" particles}
%{\bf ``Free" particle(s)}:
\ssec{Scattering States}
%\emph{Scattering states}
The scattering states may be studied here using standard non-relativistic quantum theory. The consequences of an incoming plane wave with wavenumber, $k=\sqrt{2\m E}$, scattered by a spherically symmetric potential may be calculated by writing the full wavefunction, up to a normalization constant, as
\begin{equation}
\Psi=\sum_{\ell=0}^\infty\[ j_\ell(kr)  + ik a_\ell h^{(1)}_\ell(kr) \]   (2\ell+1) i^\ell P_\ell(\cos{\th})\,,
\end{equation}
and using the standard relation between the scattering coefficients, $a_\ell$, and phase shifts, $\de_\ell$,
\begin{equation}\label{3d_aell_to_phase_shift_relation}
a_\ell=\f{1}{k}e^{i\de_\ell}\sin{\de_\ell}\,.
\end{equation}

Beginning with $\ell=0$, a perturbative expansion of \eqref{spher_symm_SAE_condition} is performed that gives, to zeroth order in $kr_\star$,
\begin{equation}\label{scattering_eqn}
1+a_0\(ik +\f{1}{r_\star} -\(\f{1}{2} k^2 + \f{1}{r_\star^2}\) Z_0(r_\star)
\)=0\,.
\end{equation}
Solving for $a_0$ and requiring that it be independent of $r_\star$ results in a linear differential equation for $Z_0$:
\begin{equation}
Z_0'(r)= \f{4 Z_0(r)-2r}{r\(2+(kr)^2\)}\,.
\end{equation}
The full solution to this equation may be obtained, but only its expansion up to order $r_\star^2$ is needed to satisfy \eqref{spher_symm_SAE_condition}, therefore
\begin{equation}\label{ell=0_Z_soln}
Z_0(r_\star)= r_\star - \f{r_\star^2}{ L}\,,
\end{equation}
where $L$ is an arbitrary integration constant whose form is chosen for later convenience. From \eqref{scattering_eqn} it follows that
\begin{equation}
a_0=\f{i L}{k L-i}
\end{equation}
or
\begin{equation}
k\cot{\de_0}= -\f{1}{L}\,,
\end{equation}
where $L$ is referred to as the scattering length, consistent with standard notation (e.g. \cite{weinberg2012lectures}).

It becomes more cumbersome to use this method for the $\ell\neq0$ cases; instead a power series for $Z_\ell(r)$ is postulated and the coefficients are determined from \eqref{spher_symm_SAE_condition}, expanded to zeroth order in $kr_\star$. Making the ansatze
\begin{equation}\label{FreeZansatze}
Z_\ell(r)=\sum_{n=0} A_n r^n\,,
\end{equation}
the first couple $\ell\neq0$ results are found to be
\begin{align}\label{freeparticleZ_lneq0}
Z_1(r)&=\f{r}{2}+ \f{k^2r^3}{4}\notag\\
Z_2(r)&=\f{r}{3}+ \f{k^2r^3}{27}\,.
%Z_3(r)&=\f{r}{4}+ \f{k^2r^3}{80} + \f{7k^4 r^5}{4800}\,,%+\f{k^4r^5}{243}
\end{align}

%There are two distinct problems with the $\ell\neq0$ cases, however:
The problem with \eqref{freeparticleZ_lneq0} is that the $Z_{\ell\neq0}$ depend on $k$, meaning that modes of different energy satisfy different boundary conditions. In this case the Hamiltonian would fail to be exactly Hermitian, and therefore the theory fails to be exactly unitary. The only way to avoid this would be to impose the \emph{additional} condition $f_{\ell\neq0}'(r_\star)=0$. However, since equation \eqref{spher_symm_SAE_condition} implies that $f_{\ell\neq0}(r_\star)$ must also vanish,  these two conditions cannot both be demanded without losing self-adjointness.

%since these two conditions should hold separately would imply that the system would be over-specified. For $r_\star\neq0$, if there are any systems that can obey both of these conditions, they are highly contrived and could occur only for special values of $r_\star$ and a discrete set of $k$. The number of such systems would form a set of measure zero so that, in effect, the condition is never satisfied.

The only way out of this is to take the limit $r_\star\to0$. However, in the limit $r\to0$, $f_\ell(r)\propto r^{-(\ell+1)}$ if $a_{\ell\neq0}\neq0$; this behavior is too singular for the wavefunction to be normalized. The only way out of this conundrum is to have $a_{\ell\neq0}=0$.

To summarize, for the method described here to work requires both
\begin{equation}
r_\star\to0
\end{equation}
and
\begin{equation}
a_{\ell\neq0}=0\,.
\end{equation}
%from which it follows that both $f_\ell(r)$ and $f_\ell'(r)$ vanish in the $r\to0$ limit.
Since $a_0$ is the only non-zero scattering coefficient, the standard relation between cross section, $\s$, and phase shifts then gives
\begin{equation}
\s= 4\p L^2\,,
\end{equation}
a standard textbook result for the scattering from, e.g. a hard sphere of radius $L$.

%\emph{Bound States}, 
\ssec{Bound States}
If bound states exist for $\ell=0$, they have the form
\begin{equation}
\psi=\f{e^{-\k r}}{r}\,,
\end{equation}
where $\k^2\equiv -2\m E$. A perturbative expansion of \eqref{spher_symm_SAE_condition} for $\ell=0$ yields, to lowest order,
\begin{equation}
1-Z_0(r_\star)\(\k+\f{1}{r_\star}\)=0\,.
\end{equation}
One can solve for $\k$ and require independence of $r_\star$ to find $Z_0(r_\star)$; the solution is identical to \eqref{ell=0_Z_soln}. Therefore, it follows that
\begin{equation}\label{free_particle_kappa}
\k=\f{1}{L}\,,
\end{equation}
which requires $L>0$ if there is to be a bound state. In that case, temporarily putting back factors of  $\hbar$, the bound (ground) state energy is
\begin{equation}\label{anomalous_bound_energy}
E_0=-\f{\hbar^2}{2\m  L^2}\,.
\end{equation}

That an anomalous bound state can appear should not be surprising in light of many known systems with short-range interactions. Consider the deuteron, a bound state of a proton and neutron, with a binding energy of approximately $2.2\, \MeV$. In that state $\m\simeq m_n/2\simeq470\,\MeV$, which indicates $L\simeq 4.3\,\text{fm}$, consistent with known results \cite{Bethe:1949yr}.  That $L$ is bigger than the proton radius (the natural cutoff)  by a factor of 5 further supports the method advocated here. Other systems that appear to admit description by this method include the dineutron resonance (see \cite{Spyrou:2012zz, Marques:2012nc,Hagino:2013zca}), as well as halo-dimers found in Feshbach-resonant systems (see e.g \cite{chin2010feshbach}). 

In addition to the obvious breaking of translation invariance, it is worth noting the anomalous breaking of scale invariance in this system\footnote{Despite the existence of the mass, $\m$, there is no intrinsic length scale here because nowhere does $c$ appear in the theory.}; i.e., there is a characteristic scale that appears in the solutions even though there is none in the Hamiltonian.

\sec{The isotropic harmonic oscillator} 
%\emph{\bf The isotropic harmonic oscillator:} 
The potential here is $V(r)=\f{1}{2}\m\o^2 r^2$. Similar to the approach for the Coulomb problem \cite{Beck:inprep}, it is best to choose the two linearly-independent solutions to the Schrodinger equation according to their large $r$ behavior so that one solution may be immediately discarded since it is not normalizable. 

Without loss of generality the radial solutions may be written
\begin{align}\label{3d_HO_soln}
f(r)\!&=\!{\cal N}_S\, e^{-1/2\,\(Kr\)^2}\!\(Kr\)^\ell\! \times\!\,U\(\f{3+2\ell-2\vep}{4}\Bigg|\f{1}{2}\Bigg| \(Kr\)^2\),
\end{align}
where $K\equiv \sqrt{\m\o}$, $\vep\equiv E/\o$, and $U$ is Kummer's confluent hypergeometric function. 
Here, the expansion of \eqref{spher_symm_SAE_condition} to zeroth order in $Kr_\star$, with the power-law ansatze \eqref{FreeZansatze} for $Z_\ell$, shows that $Z_0(r_\star)$ is identical in form to \eqref{ell=0_Z_soln}, while the first two $Z_{\ell\neq0}$ are
\begin{align}\label{HOZ_lneq0}
Z_1(r)&=\f{r}{2}+ \f{\varepsilon K^2r^3}{2}\notag\\
Z_2(r)&=\f{r}{3}+ \f{2\varepsilon K^2r^3}{27}\,.%+\f{k^4r^5}{243}
\end{align}
Identifying $2\varepsilon K^2=2E\m$ with the particle's momentum-squared, one can see that the first few $Z_\ell$ are identical in form to those of the free particle; heuristically, this is because particles are essentially free in the $r\ll K^{-1}$ limit.
%\flag{Since the approximation in this analysis is $Z_\ell=Z_0$}, the only way to satisfy \eqref{spher_symm_SAE_condition} for small and arbitrary $K r_\star$ is to have the 

As in the free particle case, there is a problem with the $\ell\neq0$ states unless $f_\ell'(r)$ vanishes and $f_\ell(r)$ is normalizeable in the $r_\star\to0$ limit. The only way to satisfy these conditions is to have the canonical spectrum for the $\ell\neq0$ states, namely
\begin{align}
\varepsilon&=2n_r + \ell +\f{3}{2}\notag\\
&\equiv n +\f{3}{2}~~~~~~~~\text{(for $\ell\neq0$)}\,,
\end{align}
where $n=0,1,2,\dots$.

For $\ell=0$, however, the perturbative expansion of \eqref{spher_symm_SAE_condition} with $Z_0$ given by \eqref{ell=0_Z_soln} reveals that
\begin{equation}\label{3d_HO_trans_eqn}
2K L=\f{\Gamma(\f{1-2\varepsilon}{4})}{\Gamma(\f{3-2\varepsilon}{4})}~~~~~~~~\text{(for $\ell=0$)}\,.
\end{equation}
Equation \eqref{3d_HO_trans_eqn} is a transcendental relation for the allowable values of $\varepsilon$; the right hand side of \eqref{3d_HO_trans_eqn} is plotted in Figure 1 along with lines of fixed $2K L$;  however there are some limiting cases that have analytic forms:\\
\\
{\bf (1)} $\abs{KL}\gg 1 $:  Here the modified spectrum looks like a perturbation of the canonical odd-$n$ spectrum,
\begin{equation}\label{3dHO_perturbed_odd_spectrum}
\varepsilon_{\ell=0}\simeq 2n+\f{1}{2} -\f{1}{\p\sqrt{n}}\,\f{1}{KL}\,,
\end{equation}
which, according to \eqref{spher_symm_SAE_condition} and \eqref{ell=0_Z_soln}, is consistent with the fact that $\Psi(r_\star)\to0$ in the limit $\abs{L}\to \infty$. Another qualitative difference from the canonical case is the anomalous state for $L>0$, addressed below.\\
\\
{\bf (2)} $\abs{KL}\ll 1$:  Here, the modified spectrum is a perturbation to the canonical even-$n$ spectrum; this is consistent with the fact that $\Psi'(r_\star)\to0$ in the limit $L\to 0$. However, the shift in those energy levels \emph{grows} as $\sqrt{n}$, and that perturbative analysis eventually breaks down at very large $n$. It is then better to expand around the higher canonical odd levels, as in \eqref{3dHO_perturbed_odd_spectrum}. In summary, it can be checked that
\begin{equation}
\varepsilon\simeq
\begin{cases}
2n+\f{3}{2} + \f{4\sqrt{n}}{\p}K L,~~~~~&n  \ll (K L)^{-2}\\
2n+\f{5}{2} - \f{1}{\p\sqrt{n}K L}, &n  \gg (K L)^{-2}\,.
\end{cases}
\end{equation}
{\bf (3)} $L>0$: As seen in Figure 1, for all $L>0$ the ground state is anomalous. Specifically, when $0<L\ll K^{-1 }$ it may be checked that $\abs{E_0}\gg\o$ and to lowest order it is independent of $K$, having the same form as \eqref{anomalous_bound_energy}. It is noteworthy, but not surprising that this result is identical to the free particle; the state is for all intents and purposes localized to the region $x\lesssim L$, so that the ratio of the average potential to bound state energy is roughly ${\cal O}(KL)^4\ll1$, meaning that the particle is essentially free.

These results are not entirely new. The authors in Ref. \cite{Busch1998} considered the physics of two cold atoms in a harmonic trap subject to a contact interaction by explicitly including a regularized delta function potential in the Hamiltonian. In that work the authors derive a version of equation \eqref{3d_HO_trans_eqn} for the energy levels corresponding to the relative motion of the atoms. Here, those results been derived from a different perspective and, arguably, in a more physically transparent and mathematically simple way.
\begin{figure}
  \begin{center}
    \includegraphics[scale=.45]{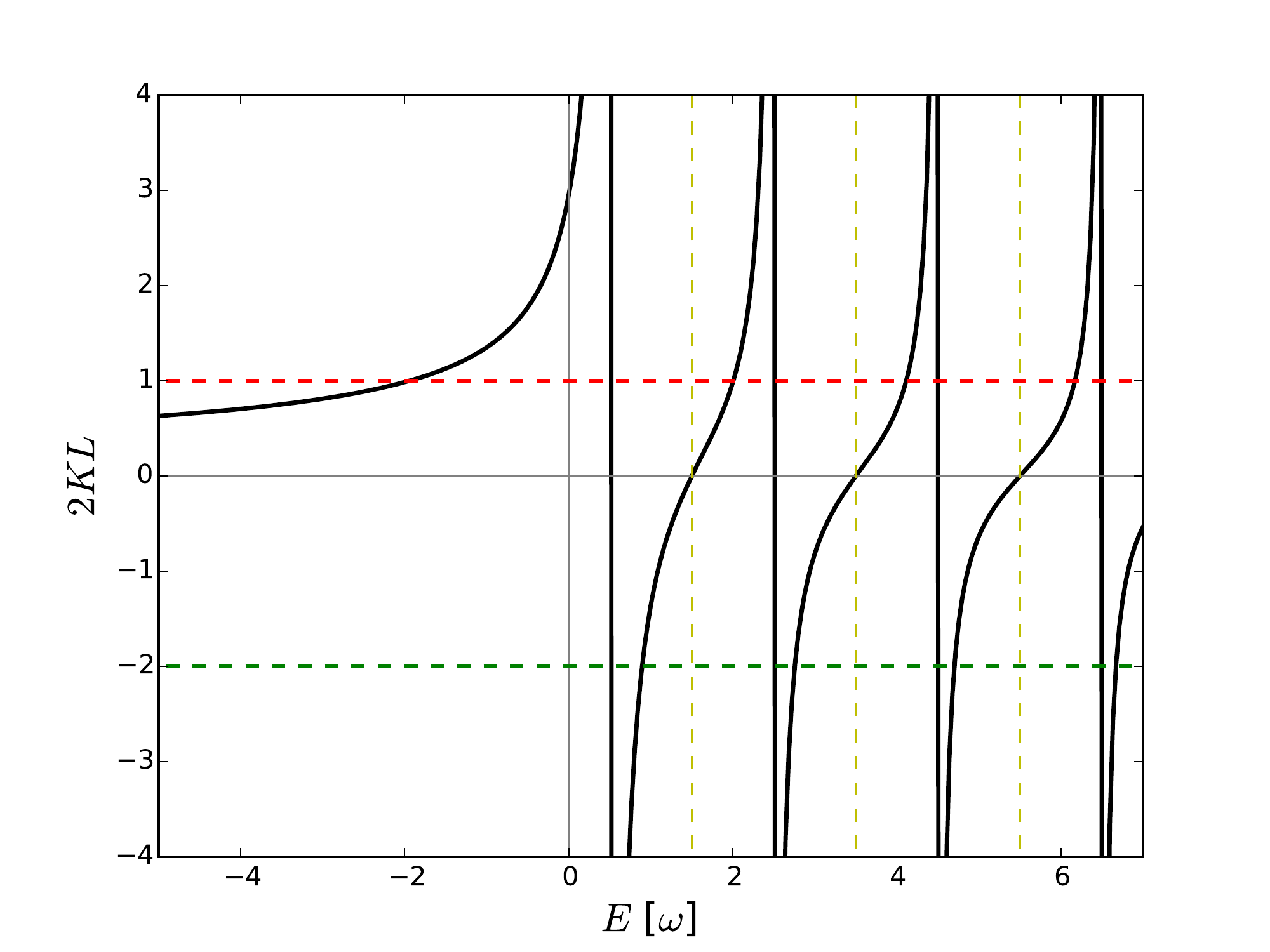}\caption{\emph{Modified harmonic oscillator spectrum.} The right hand side of \eqref{3d_HO_trans_eqn} is plotted in black and each horizontal dashed line are examples that correspond to a different theory with a fixed $L$. The energy spectrum corresponds to the intersection points.}
  \end{center}
\label{Fig_3d_HO}
\end{figure}

Concerning the symmetry properties of this system, it is well-known that the canonical harmonic oscillator in $N$ dimensions is symmetric under unitary transformations amongst the lowering and raising operators, $a_i$ and $a^\dagger_i$ -- that is to say it is symmetric under the group $U(N)$. But here this symmetry is anomalously broken and, likewise, the standard operator method fails for all boundary conditions except for the canonical one, namely $L=0$. The $U(3)$ symmetry is consequently anomalously broken to the spherical symmetry group, $SO(3)$.

To more easily illustrate this last point, consider the analogous problem on the half line in one dimension. The  operator method uses the decomposition of the Hamiltonian in terms of the lowering and raising operators, $a$ and $a^\dagger$.  These differential operators obey the standard commutation relations, so it is straightforward to show that if a state $\Psi$ is an eigenfunction of the Hamiltonian with energy $E$, then $H$ acting on the function $a^\dagger \Psi$, for example, returns the value $(E+\o)$ times $a^\dagger \Psi$. It is a function, not a state, because the objects $a^\dagger \Psi$ and $a \Psi$ do not generally lie in the domain of $H$, i.e. they do not generally obey the same boundary conditions as $\Psi$.  As a corollary, if the lowering operator, $a$, is  applied to the ground state wavefunction  it only vanishes for the special case that $L\to0$. The $U(1)$ symmetry of the one-dimensional Hamiltonian,  $a\to e^{i\th}a$, is only superficial -- it is not an actual symmetry of the system\footnote{A similar effect is found for the harmonic oscillator on a conical space \cite{AlHashimi:2007ci}.}.
 
% The argument is similar in three dimensions wherein it is standard to assume separation of variables into cartesian coordinates. However, the leading order behavior of the solution \eqref{3d_HO_soln} is generally logarithmic in $r$, so this function cannot be separated into products of functions of the separate cartesian coordinates. 

\sec{The Coulomb potential}
%{\bf The $1/r$ potential:}
Here some of the results of \cite{Beck:inprep} are advertised, however, they are derived using the artificial boundary method. The potential is written $V(r)=Z_1Z_2 \a c/r$, where $c$ is the speed of light and $Z_{1,2}$ are integer multiples of the proton charge; hereafter $c$ is set equal to unity. It is possible to consider deviations to the canonical hydrogen spectrum as in \cite{Beck:inprep}, but for both brevity and novelty a different result is presented, namely the scattering states and anomalous bound state of a repulsive Coulomb system. 

%\emph{Scattering states:}
\ssec{Scattering states}
One may write the $\ell$-channel solutions as
\begin{multline}\label{CoulombSoln}
f_\ell(r)=e^{-ik r}\(kr\)^\ell \(\,_1F_1\[1+\ell -i\f{q}{k},2+2\ell, 2ikr\]\right.\\
\left. + B_\ell\, U\[1+\ell -i\f{q}{k},2+2\ell, 2ikr\]\)\,,
\end{multline}
where $q\equiv Z_1 Z_2 \m \a$. As is standard, the $r\to \infty$ analysis decomposes the wavefunction into incoming and outgoing waves as
\begin{equation}
\lim_{r\to\infty} f_\ell(r)\!\sim\!  \f{1}{kr}\!\(e^{i(kr - q/k \ln{2kr}+2\de_\ell)} \!-\! (-1)^\ell e^{-i(kr - q/k \ln{2kr})}\)\,.
\end{equation}
Using the form of solution given in \eqref{CoulombSoln}, the full phase factor may be written as
\begin{equation}
e^{2i\de_\ell}=e^{2i\bar{\de}_\ell}\(1 -(-1)^\ell B_\ell e^{-\p q/k} \f{\Gamma(1+\ell+i\f{q}{k})}{(2\ell+1)!}  \)^{-1}\,,
\end{equation}
where the canonical phase factor,
\begin{equation}
e^{2i\bar{\de}_\ell}=\f{\Gamma(1+\ell+i\f{q}{k})}{\Gamma(1+\ell-i\f{q}{k})}\,.
\end{equation}
One could take the same approach as for the free particle by taking the $kr_\star\to0$ limit of \eqref{spher_symm_SAE_condition} and obtaining a differential equation for $Z_\ell$ by requiring $B_\ell$ to be independent of $r_\star$. Instead, similar to the above analysis, one can simply make the ansatze\footnote{The coefficient of $r$ in the argument of the logarithm of \eqref{CoulombZAnsatze} has been chosen as $\abs{q}$ since this appears to be the most natural scale to use. Although this choice it is not unique, any deviation from it can be absorbed by a redefinition of $A_n$. Furthermore, in the cases of interest presented here, this would require only a small redefinition of $A_n$.}
\begin{equation}\label{CoulombZAnsatze}
Z_\ell(r)=\sum_{n=0} r^n\(A_n + B_n \ln{\abs{q}r}\)
\end{equation}
and determine the coefficients $A_n$ and $B_n$ that ensure that \eqref{spher_symm_SAE_condition} holds in the limit $kr_\star\to0$. It may be checked that
\begin{equation}
Z_0(r)\simeq r+ \(-\f{1}{L}+ 2q\ln{\abs{q}r}\) \! r^2\label{Coulomb_Z_0}\,,
\end{equation}
while
\begin{align}\label{Coulomb_Zs}
Z_1(r)&\simeq \f{r}{2} -\f{qr^2}{4}+ \f{\(2k^2+3q^2\)r^3}{8}\\
Z_2(r)&\simeq \f{r}{3} -\f{qr^2}{18}+ \f{\(2k^2+q^2\)r^3}{54} - \f{q\(10k^2+3q^2\) r^4}{324}\,.
\end{align}
(Note that these results reduce to those of the free particle when $q\to0$.)    As above, it must be that both
%be that $f_{\ell\neq0}'(r_\star)=0$ to ensure the Hamiltonian is strictly Hermitian, and that the wavefunctions be normalizeable in the  limit; both of these conditions require 
$r_\star\to0$ and $B_{\ell\neq0}=0$ to ensure that both the Hamiltonian is strictly Hermitian, and that the wavefunctions can be normalized. The only non-zero coefficient is found to be
\begin{align}\label{B_0}
B_0&=-\f{2qL\,\Gamma(-i\f{q}{k})}{1-ikL+qL\(4\g  +2\ln{2i\f{k}{\abs{q}}}+2 \,\psi(1-i\f{q}{k})\)}\,,
\end{align}
where $\psi(x)=\Gamma'(x)/\Gamma(x)$ is the digamma function. As cumbersome as \eqref{B_0} may appear, some considerable simplifications can be made if $L\ll \abs{q}^{-1}$, which is true in the cases of interest. The following approximation can then be made:
\begin{equation}
B_0\simeq-\f{2qL\,\Gamma(-i\f{q}{k})}{1-ikL}~~~~~~~\text{(for $L\ll \abs{q}^{-1}$)}\,,
\end{equation}
and therefore
\begin{align}
e^{2i(\de_\ell-\bar{\de_\ell})}&\simeq \f{i+kL}{i+kL +2\p q L-2\p qL\coth{\p q/k}}\notag\\
&\sim\f{i+kL}{i-kL}\,,~~~~~~~~~~\text{(for $k\gg q$)}\,.
\end{align}
This reveals that a resonance exists at $k\simeq L^{-1}$, which coincides exactly with the free particle results when $q\to0$. It is worth noting that the above analysis is applicable for either sign of $q$, so these scattering results apply equally well for repulsive or attractive Coulomb potentials.

%\emph{Bound States:} 
\ssec{Bound States}
Requiring normalizability as $r\to\infty$, it is best to write the radial solutions of this system as
\begin{equation}
%f(r)=e^{-\k r}  \(\m \a r\)^\ell U\,\(1+\ell+\f{\m \a}{\k}\Big|2\ell\Big|2\k r\)\,,
f(r)=e^{-\k r}  \(\abs{q} r\)^\ell U\,\(1+\ell+\f{q}{\k}\Big|2\ell\Big|2\k r\)\,,
\end{equation}
where $\k^2\equiv-2\m E_0$. Since \eqref{spher_symm_SAE_condition} should hold for arbitrarily small $r_\star$, a first-order differential equation for $Z_0$ may be obtained and solved perturbatively as above; however, we already have the result for $Z_0$ given in \eqref{Coulomb_Z_0}.
%For brevity, only the solution is presented:
%\begin{equation}\label{Coulomb_Z_0}
%Z_0(r_\star)\simeq r_\star+ \(-\f{1}{L}+ 2q\ln{q r_\star}\) \! r_\star^2\,.
%\end{equation}
%The coefficient of $r_\star$ in the argument of the logarithm has been chosen as $q$ since this appears to be the most natural scale to use. Although this choice it is not unique, any deviation from it can be absorbed by a redefinition of $L$;  the correction, which is of the order $q^{-1}$, is generally quite small in the cases of interest.

Using \eqref{Coulomb_Z_0} in \eqref{spher_symm_SAE_condition}, the spectrum is determined by the transcendental equation
\begin{equation}\label{Coulomb_energy_eqn}
q L =\(\f{\k}{q} -4\g -2\ln{\f{2\k}{\abs{q}}} - 2\psi\(1+\f{q}{\k}\)\)^{-1}\,,
\end{equation}
a result that was first found (for the attractive potential) in \cite{Beck:inprep}. Assuming a repulsive potential ($q>0$), a plot of the right hand side of \eqref{Coulomb_energy_eqn} is given in Figure 2 along with lines of fixed $q L$. If $0< L\ll q^{-1}$, then $\k$ is approximately independent of $q$ and given by \eqref{free_particle_kappa}, and therefore $E_0$ is given by \eqref{anomalous_bound_energy}. 

\begin{figure}
  \begin{center}
    \includegraphics[scale=.45]{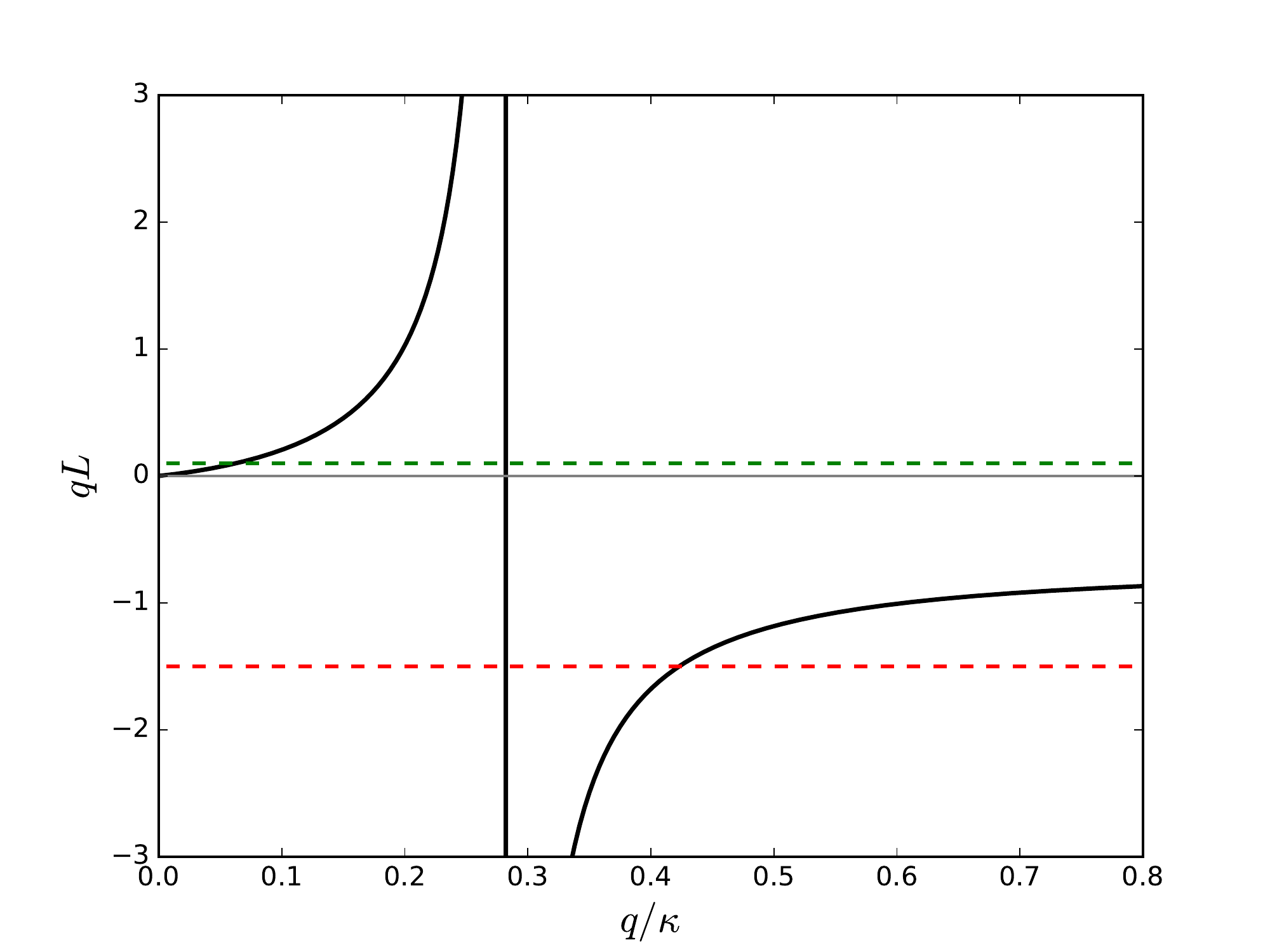}\caption{\emph{Possible bound state in a repulsive Coulomb potential}. The right side of \eqref{Coulomb_energy_eqn} for $q>0$ is plotted in black. Horizontal dashed lines are examples that correspond to different theories with fixed value of $L$; the bound state energy corresponds to an intersection point.}
  \end{center}
\label{Fig_3d_Coulomb}
\end{figure}

One physical application for $L>0$ is apparently the bound state between a deuteron and proton, namely a $^3\text{He}$ nucleus. The proton separation energy in that system is measured to be roughly $5.5\,\MeV$ \cite{NNDC} and, given that the reduced mass is $\m\simeq 625\,\MeV$, to fit this system would require $L\simeq2.4\, \text{fm}$; the fact that $L^{-1}$ here is roughly twice the value of that for the deuteron appears reasonable.   Equation \eqref{Coulomb_energy_eqn} also indicates that a bound state is possible for $q L<-1/\ln{(2)}$, but whether those states are ever realized in nature is obscure to the author.

Lastly, since the phenomena here depend on $\ell$ necessarily means that the celebrated symmetry of the $1/r$ potential -- actually, $SO(3,1)$ here -- is anomalously broken to $SO(3)$, a well-known result.

\sec{Discussion}
%{\bf Discussion}:
 The method described here captures short-distance physical effects in an effective way by imposing an artificial boundary of ignorance, beyond which the theory is not specified. The set of allowable boundary conditions may be viewed as a way to describe the set of possible UV completions of the theory. The method was applied to the non-relativistic free particle, isotropic harmonic oscillator, and the Coulomb potential, all in three spatial dimensions. Well-known non-trivial results were derived, but in a way that did not require additional short-distance physics to be added to the theory \emph{a priori}, for example, by the introduction of $\de$-function potentials. %Analogous effects to those presented here in one- and two-dimensional systems and in relativistic systems will be addressed elsewhere \cite{Jacobs2016}.

An analysis of this type is required to model \emph{any} physical system unless there is some prior knowledge, e.g. an expected symmetry that restricts the possible short-distance phenomena present.  In the Coulomb system, short-ranged fine structure effects can be described in terms of a boundary condition with a natural magnitude of the order $\a^2\sim 10^{-4}$, but it would be very interesting to look for analogous corrections in strongly coupled systems where, naively, the relevant coupling is expected to be of order $\a_{s}^2\sim {\cal O}(1)$.

The method presented here may also offer advantages over other approaches that use delta function potentials, as those distributions are ill-suited to describe anything besides structureless, point-like interactions with a high degree of symmetry. It would be interesting to consistently extend this approach in a way that does not require taking the region of ignorance to zero size, thereby providing a description of interactions that take place over a finite range. This would likely require relaxing the standard unitarity requirement so that probability current can flow back and forth through the artificial boundary at $r_\star$, and this is the subject of an on-going investigation \cite{Jacobs2016}.
\\
\\
\\
\emph{Acknowledgements --} I am indebted to Harsh Mathur for introducing me to the subject of self-adjoint extensions in quantum mechanics, for emphasizing their application to contact interactions, and for many useful conversations and helpful insights. I have also benefited greatly from conversations with Claudia de Rham and Andrew Tolley on the connection between this work and effective field theories with boundaries. I also thank Glenn Starkman, Scott Beck, Craig Copi, Andrew Matas, Lorenzo Reverberi, Will Horowitz, and Chris Clarkson for stimulating discussions. I acknowledge support from the Claude Leon Foundation and also thank the physics department of Case Western Reserve University for hospitality during stages of this work. 
\bibliographystyle{apsrev}
\bibliography{EQM_bib}

\begin{thebibliography}{20}
\expandafter\ifx\csname natexlab\endcsname\relax\def\natexlab#1{#1}\fi
\expandafter\ifx\csname bibnamefont\endcsname\relax
  \def\bibnamefont#1{#1}\fi
\expandafter\ifx\csname bibfnamefont\endcsname\relax
  \def\bibfnamefont#1{#1}\fi
\expandafter\ifx\csname citenamefont\endcsname\relax
  \def\citenamefont#1{#1}\fi
\expandafter\ifx\csname url\endcsname\relax
  \def\url#1{\texttt{#1}}\fi
\expandafter\ifx\csname urlprefix\endcsname\relax\def\urlprefix{URL }\fi
\providecommand{\bibinfo}[2]{#2}
\providecommand{\eprint}[2][]{\url{#2}}

\bibitem[{\citenamefont{Bonneau et~al.}(2001)\citenamefont{Bonneau, Faraut, and
  Valent}}]{Bonneau:1999zq}
\bibinfo{author}{\bibfnamefont{G.}~\bibnamefont{Bonneau}},
  \bibinfo{author}{\bibfnamefont{J.}~\bibnamefont{Faraut}}, \bibnamefont{and}
  \bibinfo{author}{\bibfnamefont{G.}~\bibnamefont{Valent}},
  \bibinfo{journal}{Am.J.Phys.} \textbf{\bibinfo{volume}{69}},
  \bibinfo{pages}{322} (\bibinfo{year}{2001}), \eprint{quant-ph/0103153}.

\bibitem[{\citenamefont{Essin and Griffiths}(2006)}]{essin2006quantum}
\bibinfo{author}{\bibfnamefont{A.~M.} \bibnamefont{Essin}} \bibnamefont{and}
  \bibinfo{author}{\bibfnamefont{D.~J.} \bibnamefont{Griffiths}},
  \bibinfo{journal}{Am. J. Phys.} \textbf{\bibinfo{volume}{74}},
  \bibinfo{pages}{109} (\bibinfo{year}{2006}).

\bibitem[{\citenamefont{Dasarathy et~al.}(2016)\citenamefont{Dasarathy,
  Jones-Smith, Kerr, and Mathur}}]{Dasarathy:unpub}
\bibinfo{author}{\bibfnamefont{A.}~\bibnamefont{Dasarathy}},
  \bibinfo{author}{\bibfnamefont{K.}~\bibnamefont{Jones-Smith}},
  \bibinfo{author}{\bibfnamefont{A.}~\bibnamefont{Kerr}}, \bibnamefont{and}
  \bibinfo{author}{\bibfnamefont{H.}~\bibnamefont{Mathur}},
  \bibinfo{journal}{(unpublished)}  (\bibinfo{year}{2016}).

\bibitem[{\citenamefont{Jackiw}(1995)}]{jackiw1995diverse}
\bibinfo{author}{\bibfnamefont{R.}~\bibnamefont{Jackiw}},
  \emph{\bibinfo{title}{{Diverse topics in theoretical and mathematical
  physics}}} (\bibinfo{year}{1995}).

\bibitem[{\citenamefont{Coutinho et~al.}(1997)\citenamefont{Coutinho, Nogami,
  and Perez}}]{Coutinho}
\bibinfo{author}{\bibfnamefont{F.}~\bibnamefont{Coutinho}},
  \bibinfo{author}{\bibfnamefont{Y.}~\bibnamefont{Nogami}}, \bibnamefont{and}
  \bibinfo{author}{\bibfnamefont{J.~F.} \bibnamefont{Perez}},
  \bibinfo{journal}{Journal of Physics A: Mathematical and General}
  \textbf{\bibinfo{volume}{30}}, \bibinfo{pages}{3937} (\bibinfo{year}{1997}).

\bibitem[{\citenamefont{Roy and Stone}(2010)}]{Roy:2009vc}
\bibinfo{author}{\bibfnamefont{A.}~\bibnamefont{Roy}} \bibnamefont{and}
  \bibinfo{author}{\bibfnamefont{M.}~\bibnamefont{Stone}}, \bibinfo{journal}{J.
  Phys.} \textbf{\bibinfo{volume}{A43}}, \bibinfo{pages}{015203}
  (\bibinfo{year}{2010}), \eprint{0909.1569}.

\bibitem[{\citenamefont{Albeverio et~al.}(2012)\citenamefont{Albeverio,
  Gesztesy, Hoegh-Krohn, and Holden}}]{albeverio2012solvable}
\bibinfo{author}{\bibfnamefont{S.}~\bibnamefont{Albeverio}},
  \bibinfo{author}{\bibfnamefont{F.}~\bibnamefont{Gesztesy}},
  \bibinfo{author}{\bibfnamefont{R.}~\bibnamefont{Hoegh-Krohn}},
  \bibnamefont{and} \bibinfo{author}{\bibfnamefont{H.}~\bibnamefont{Holden}},
  \emph{\bibinfo{title}{Solvable models in quantum mechanics}}
  (\bibinfo{publisher}{Springer Science \& Business Media},
  \bibinfo{year}{2012}).

\bibitem[{\citenamefont{Beck et~al.}(2016)\citenamefont{Beck, Jacobs, and
  Mathur}}]{Beck:inprep}
\bibinfo{author}{\bibfnamefont{S.~J.} \bibnamefont{Beck}},
  \bibinfo{author}{\bibfnamefont{D.~M.} \bibnamefont{Jacobs}},
  \bibnamefont{and} \bibinfo{author}{\bibfnamefont{H.}~\bibnamefont{Mathur}},
  \bibinfo{journal}{(unpublished)}  (\bibinfo{year}{2016}).

\bibitem[{\citenamefont{Chin et~al.}(2010)\citenamefont{Chin, Grimm, Julienne,
  and Tiesinga}}]{chin2010feshbach}
\bibinfo{author}{\bibfnamefont{C.}~\bibnamefont{Chin}},
  \bibinfo{author}{\bibfnamefont{R.}~\bibnamefont{Grimm}},
  \bibinfo{author}{\bibfnamefont{P.}~\bibnamefont{Julienne}}, \bibnamefont{and}
  \bibinfo{author}{\bibfnamefont{E.}~\bibnamefont{Tiesinga}},
  \bibinfo{journal}{Reviews of Modern Physics} \textbf{\bibinfo{volume}{82}},
  \bibinfo{pages}{1225} (\bibinfo{year}{2010}).

\bibitem[{\citenamefont{Feynman}(1948)}]{Feynman:1948ur}
\bibinfo{author}{\bibfnamefont{R.~P.} \bibnamefont{Feynman}},
  \bibinfo{journal}{Rev. Mod. Phys.} \textbf{\bibinfo{volume}{20}},
  \bibinfo{pages}{367} (\bibinfo{year}{1948}).

\bibitem[{\citenamefont{Reed and Simon}(1980)}]{reed1980methods}
\bibinfo{author}{\bibfnamefont{M.}~\bibnamefont{Reed}} \bibnamefont{and}
  \bibinfo{author}{\bibfnamefont{B.}~\bibnamefont{Simon}},
  \emph{\bibinfo{title}{Methods of modern mathematical physics: Functional
  analysis, vol. 1}} (\bibinfo{publisher}{Academic Press},
  \bibinfo{year}{1980}).

\bibitem[{\citenamefont{Weinberg}(2012)}]{weinberg2012lectures}
\bibinfo{author}{\bibfnamefont{S.}~\bibnamefont{Weinberg}},
  \emph{\bibinfo{title}{Lectures on quantum mechanics}}
  (\bibinfo{publisher}{Cambridge University Press}, \bibinfo{year}{2012}).

\bibitem[{\citenamefont{Bethe}(1949)}]{Bethe:1949yr}
\bibinfo{author}{\bibfnamefont{H.~A.} \bibnamefont{Bethe}},
  \bibinfo{journal}{Phys. Rev.} \textbf{\bibinfo{volume}{76}},
  \bibinfo{pages}{38} (\bibinfo{year}{1949}).

\bibitem[{\citenamefont{Spyrou et~al.}(2012)}]{Spyrou:2012zz}
\bibinfo{author}{\bibfnamefont{A.}~\bibnamefont{Spyrou}} \bibnamefont{et~al.},
  \bibinfo{journal}{Phys. Rev. Lett.} \textbf{\bibinfo{volume}{108}},
  \bibinfo{pages}{102501} (\bibinfo{year}{2012}).

\bibitem[{\citenamefont{Marques et~al.}(2012)\citenamefont{Marques, Orr,
  Achouri, Delaunay, and Gibelin}}]{Marques:2012nc}
\bibinfo{author}{\bibfnamefont{F.~M.} \bibnamefont{Marques}},
  \bibinfo{author}{\bibfnamefont{N.~A.} \bibnamefont{Orr}},
  \bibinfo{author}{\bibfnamefont{N.~L.} \bibnamefont{Achouri}},
  \bibinfo{author}{\bibfnamefont{F.}~\bibnamefont{Delaunay}}, \bibnamefont{and}
  \bibinfo{author}{\bibfnamefont{J.}~\bibnamefont{Gibelin}},
  \bibinfo{journal}{Phys. Rev. Lett.} \textbf{\bibinfo{volume}{109}},
  \bibinfo{pages}{239201} (\bibinfo{year}{2012}), \eprint{1204.5946}.

\bibitem[{\citenamefont{Hagino and Sagawa}(2014)}]{Hagino:2013zca}
\bibinfo{author}{\bibfnamefont{K.}~\bibnamefont{Hagino}} \bibnamefont{and}
  \bibinfo{author}{\bibfnamefont{H.}~\bibnamefont{Sagawa}},
  \bibinfo{journal}{Phys. Rev.} \textbf{\bibinfo{volume}{C89}},
  \bibinfo{pages}{014331} (\bibinfo{year}{2014}), \eprint{1307.5502}.

\bibitem[{\citenamefont{Busch et~al.}(1998)}]{Busch1998}
\bibinfo{author}{\bibfnamefont{T.}~\bibnamefont{Busch}} \bibnamefont{et~al.},
  \bibinfo{journal}{Foundations of Physics} \textbf{\bibinfo{volume}{28}},
  \bibinfo{pages}{549} (\bibinfo{year}{1998}).

\bibitem[{\citenamefont{Al-Hashimi and Wiese}(2008)}]{AlHashimi:2007ci}
\bibinfo{author}{\bibfnamefont{M.~H.} \bibnamefont{Al-Hashimi}}
  \bibnamefont{and} \bibinfo{author}{\bibfnamefont{U.~J.} \bibnamefont{Wiese}},
  \bibinfo{journal}{Annals of Physics} \textbf{\bibinfo{volume}{323}},
  \bibinfo{pages}{82} (\bibinfo{year}{2008}), \eprint{0707.4379}.

\bibitem[{NND(2015)}]{NNDC}
\emph{\bibinfo{title}{National nuclear data center}} (\bibinfo{year}{2015}),
  \urlprefix\url{http://www.nndc.bnl.gov/chart/}.

\bibitem[{\citenamefont{Jacobs}(2016)}]{Jacobs2016}
\bibinfo{author}{\bibfnamefont{D.~M.} \bibnamefont{Jacobs}},
  \bibinfo{journal}{(unpublished)}  (\bibinfo{year}{2016}).

\end{thebibliography}
\end{document}